\documentclass[aps,preprint]{revtex4}%
\usepackage{amsfonts}
\usepackage{amsmath}
\usepackage{amssymb}
\usepackage{graphicx}%
\begin{document}
\title{\textbf{Magnetism via superconductivity in SF proximity structures}}
\author{\textbf{V. N. Krivoruchko, V. N. Varyukhin}}
\affiliation{Donetsk Physics \& Technology Institute NAS of Ukraine, Str. R. Luxemburg 72,
83114 Donetsk, Ukraine}
\affiliation{E-mail: krivoruc@krivoruc.fti.ac.donetsk.ua}
\date{}

\begin{abstract}
We consider the proximity effects in hybrid superconductor (S) - ferromagnet
(F) structures drawing attention to the induced ferromagnetism of the S metal.
The analysis is based on a quasiclassical theory of proximity effect for
metals in the dirty limit conditions. It is shown that, below the
superconducting critical temperature, ferromagnetic correlations extend a
distance of order of the superconducting coherence length $\xi_{S}$\ into the
superconductor, being dependent on the S/F interface parameters. We argue that
the properties of mesoscopic SF hybrids may drastically depend upon the
magnetic proximity effect, and recent experiments lend support to the model of
SF structures where the superconducting and magnetic parameters are tightly coupled.

\end{abstract}
\maketitle

\section{Introduction}

Proximity effects are phenomena stipulated by a 'penetration' of an order
parameter (of some state) from one material into another, which does not
possess such type of the order itself, due to the materials being in contact.
The leakage of superconducting correlations into a non-superconducting
material is an example of the superconducting proximity effect. For a
nonmagnetic normal metal (N) in contact with a superconductor (S), the
proximity effect has been intensively studied and well understood many years
ago [1,2]. However, in the case of SN structures we deal with a single type of
order - superconductivity. When a normal nonmagnetic metal is replaced by a
ferromagnet (F), the physics of proximity effect is much more interesting and
rich [2-20]. There are two competing states with different order parameters:
superconductivity and ferromagnetism. Due to the difference in energy between
spin-up and spin-down electrons and holes under the exchange field of a
ferromagnet, a singlet Cooper pair, adiabatically injected from a
superconductor into a ferromagnet, acquires a finite momentum $\Delta p\sim
H_{e}/\hslash v_{F}$ (here $H_{e}=\mu_{B}h_{F}$ is\ an extra energy caused by
the intrinsic magnetic field $h_{F}$ in ferromagnet; $v_{F}$ is the Fermi
velocity, $\hslash$ is the Planck constant, and $\mu_{B}$ is the Bohr
magneton). As a result, proximity induced superconductivity of the F layer is
spatially inhomogeneous and the order parameter contains nodes where the phase
changes by $\pi$. Particularly, transport properties of tunnel SF structures
in the $\pi$-phase\ state have turned out quite unusual. The phase shift of
$\pi$ in the ground state of the junction is formally described by the
negative critical current $J_{C}$ in the Josephson current-phase relation:
$J(\varphi)=J_{C}\sin(\varphi)$. The\ $\pi$-phase\ state of an SFS weak link
due to Cooper pair spatial oscillation was first predicted by Buzdin
\textit{et al}., [4,5].\ Experiments that have been performed by now on SFS
weak links [6-8] and SIFS\ tunnel junctions [9]\ directly prove the $\pi
$-phase superconductivity (I denotes an insulator). Planar tunneling
spectroscopy also reveals a $\pi$-phase shift in the order parameter, when
superconducting correlations coexist with ferromagnetic order [10]. The
superconducting phase was also measured directly [11] using SQUID's made of
$\pi$-junctions.

There is another interesting case of a thin F layer, $d_{F}<<\xi_{F}$, being
in contact with an S layer. As far as the thickness of the F layer, $d_{F}$,
is much less than the corresponding superconducting coherence length, $\xi
_{F}$, there is spin spiltting but there is no order parameter oscillation in
the F layer. Surprisingly, but it was recently predicted Refs. [12,13] that
for SFIFS tunnel structures with very thin F layers one can, if there is a
parallel orientation of the F layers magnetization, turn the junction into the
$\pi$-phase state with the critical current inversion; if there is an
antiparallel orientation of the F layers internal fields, one can even enhance
the tunnel current.\ It was shown in Refs. [14,15], that physics behind the
inversion and the enhancement of the supercurrent in this case differs from
that proposed by Buzdin \textit{et al.}\ 

While proximity induced superconductivity of the F metal in SF hybrid
structures has been intensively studied, much less attention has been paid to
a modification of the electron spectrum of a superconductor in a region near
the S/F interface due to a leakage of magnetic correlation into the
superconductor.\ Some feature of the induced magnetism (e.g., the
spin-splitting of the density of states) were found by numerical calculations
in Refs. [16-18]. To our knowledge, only recently the question of S metal
magnetization has been addressed in Refs. [19,20]. (Here we do not touch S/FI
systems, where FI stands for a ferromagnetic insulator (semiconductor). In
such systems conduction electrons penetrate the magnetic layer on much smaller
distances than in the case of metals and\ are totally reflected at the S/FI
boundary. The S/FI boundary being magnetically active rotates spins of
reflected electrons. This spin rotation occurs only as a result of a tunneling
by the quasiparticle into the classically forbidden region of the boundary.
Due to the spin rotation the exchange field is induced in a superconductor on
a distance of order of superconducting coherence length $\xi_{S}$ near an S/FI
surface [21-23]. However, in contrast to ferromagnetic metals, where the
proximity effect is pronounced, this effect is drastically reduced in S/FI structures.)

The investigation of a 'magnetic proximity' effect in SF nanostructures is the
purpose of this report. To tackle the physics, we consider an interesting and
practicable case of an SF structure of a massive\ superconducting and thin
ferromagnetic layers. Using a quasiclassical theory of superconductivity for
proximity coupled bilayer (Sec. II), we will show that for some limits the
problem can be solved analytically. Two limits will be discussed here: (i) a
weak and (ii) a strong proximity effect. Section III is the key one; here we
describe the examples of the magnetic proximity effect manifestation. We show
that\ due to induced magnetism of the S metal: (i) the superconducting phase
jumps at the S/F interface; as well as, there are (ii) additional suppression
of the order parameter near the S/F interface; (iii) the spin splitting of the
quasiparticle density of states (DOS); (iv) the appearance of the local bands
inside the energy gap; and, directly, in (v) induced equilibrium electronic
magnetization of the S layer that spreads over distance of the order of the
superconducting coherence length $\xi_{S}$. We also briefly discuss recent
experiments. Summarizing the results in Conclusion we draw attention to the
fact that in the general case, for proximity coupled SF hybrid structures both
phenomena - induced superconductivity of the F metal and induced magnetism of
the S metal - take place simultaneously and should be considered self-consistently.

\section{Quasiclassical theory of superconductivity of SF bilayer}

\subsection{Bilayer model}

Let us consider proximity effects in the bilayer of a massive superconducting
($d_{S}>>\xi_{S}$) and a thin ferromagnetic ($d_{F}<<\xi_{F}$) metals, with
arbitrary transparency of the S/F interface. Here $\xi_{S}=(D_{S}/2\pi
T_{C})^{1/2}$ and $\xi_{F}=(D_{F}/2H_{e})^{1/2}$ stand for the superconducting
coherence lengths, $D_{S,F}$ are the diffusion coefficients, $d_{S,F}$ are the
thicknesses of the S and F layers. (Henceforth, we have taken the system of
units with $\hbar=k_{B}=1$.) We assume the 'dirty'\ limit for both metals,
i.e., $\xi_{S,F}>>l_{S,F}$ where $l_{S,F}$ are the electron mean free paths.
It is also assumed that the superconducting critical temperature of the F
material equals zero. All quantities are assumed to depend only on a single
coordinate $x$ normal to the interface surface of the materials. We also
expect that the F layer has a homogeneous (monodomain) magnetic structure with
magnetization aligned parallel to the interface, so that there is no
spontaneous magnetic flux penetrating into the S layer. Under these
conditions, the only magnetic interaction which can affect the superconductor
is the short-range exchange interaction between the superconducting
quasiparticles and magnetic moments into the ferromagnet.

\subsection{Main equations}

As is well know, the superconductivity of 'dirty'\ metals is conveniently
described by the quasiclassical Usadel equations [24] for the normal,
$G_{\sigma\sigma^{\prime}}(x,\omega)$ and\ $\widetilde{G}_{\sigma
\sigma^{^{\prime}}}(x,\omega)$, and anomalous, $F_{\sigma\sigma^{^{\prime}}%
}(x,\omega)$ and\ $\widetilde{F}_{\sigma\sigma^{^{\prime}}}(x,\omega)$, Green
functions, integrated over energy and averaged over the Fermi surface. (Green
functions are defined in a standard way, see, e.g. Ref. [25]). It can be
shown, that for singlet pairing and in the absence of spin-flip scattering,
the whole system of Usadel equations decomposes into two equivalent subgroups,
which go over to each other under interchange of the spin indices
($\sigma=\uparrow,\downarrow$) $\uparrow\longleftrightarrow\downarrow$ and
reversal of the exchange field sign, $H_{e}\longleftrightarrow-H_{e}$

It is convenient to take into account the normalization of the Green function,
$G_{F\uparrow\uparrow}\widetilde{G}_{F\downarrow\downarrow}+F_{F\downarrow
\uparrow}F_{F\uparrow\downarrow}^{+}=1$, explicitly and to introduce [2]
modified Usadel functions $\Phi_{S}$, $\Phi_{F}$, defined by the relations
$\Phi_{S}=\omega F_{S}/G_{S}$, $\Phi_{F}=\tilde{\omega}F_{F}/G_{F}$, etc. Then
we can recast the equations for the S layer in terms of these functions. We
specialize the discussion to a geometry when all quantities depend on a single
coordinate $x$, normal to the S/F interface. For the superconducting metal we
have ($x\geq0$):
\begin{equation}
\Phi_{S}=\Delta_{S}+\xi_{S}{}^{2}\frac{\pi T_{C}}{\omega G_{S}}[G_{S}{}%
^{2}\Phi_{S}^{\prime}]^{\prime},\text{ \ \ \ \ \ }G_{S}=\frac{\omega}%
{(\omega^{2}+\Phi_{S}\tilde{\Phi}_{S})^{1/2}},
\end{equation}
with the superconducting order parameter $\Delta_{S}(x)$ determined by the
self-consistency equation:
\begin{equation}
\Delta_{S}\ln(T/T_{C})+2\pi T\sum_{\omega>0}[(\Delta_{S}-\Phi_{S}G_{S}%
)/\omega]=0,
\end{equation}
Here the prime denotes differentiation with respect to a coordinate $x$, and
in Eq. (2) the summation over frequencies is cut off by the Debye frequency
$\omega_{D}$. For the F metal we have ($-d_{F}\leq x<0$):%
\begin{equation}
\Phi_{F}=\xi^{2}\frac{\pi T_{C}}{\tilde{\omega}G_{F}}[G_{F}{}^{2}\Phi
_{F}^{\prime}]^{\prime},\text{ \ \ \ \ }G_{F}=\frac{\tilde{\omega}}%
{(\tilde{\omega}^{2}+\Phi_{F}\tilde{\Phi}_{F})^{1/2}}%
\end{equation}
Here $\widetilde{\omega}=\omega+iH_{e}$, and $\omega\equiv\omega_{n}=\pi
T(2n+1)$, $n=\pm1,\pm2,\pm3,...$ is Matsubara frequency. Assuming the symmetry
of the system with respect to the rotation in spin space both in the F and in
the S layers, we drop the spin indices, apart from the specified cases.
We\ also put for the F metal a vanishing value of the bare superconducting
order parameter $\Delta_{F}=0$, while the pair amplitude $F_{F}\neq0$ due to
proximity with the superconductor.

The equations for the functions $\tilde{\Phi}_{S}$ and $\tilde{\Phi}_{F}%
$\ have a form analogous to (1)--(3); note that $\tilde{\Phi}(\omega
,H_{e})=\Phi^{\ast}(\omega,-H_{e})$. In Eq. (3) we write our formulas for the
F metal using the effective coherence length of normal nonmagnetic (N) metal
with the diffusion coefficient $D_{F}$, $\xi=\left(  D_{F}/2\pi T_{C}\right)
^{1/2}$, instead of $\xi_{F}=\left(  D_{F}/2H_{e}\right)  ^{1/2}$, to have a
possibility to analyze both limits $H_{e}\rightarrow0$ (SN bilayer) and
$H_{e}>>\pi T_{C}$. The relation on the ferromagnetic layer thickness one may
read as $d_{F}<<\min(\xi_{F},\xi)$.

\subsection{Boundary conditions}

The Eqs. (1)-(3) should be supplemented with the boundary conditions in the
bulk of the S metal and at the external surface of the F layer. Far from the
S/F interface, $x>>\xi_{S}$, for the S layer we have the usual boundary
conditions in the bulk of the S metal: $\Phi_{S}(\infty)=\Delta_{S}%
(\infty)=\Delta_{0}(T)$, where $\Delta_{0}(T)$ is the BCS value of the order
parameter. At the external surface of the F metal $\Phi_{F}^{/}(-d_{F})=0$.
The relations at the S/F interface we obtain [26] by generalizing the results
of Kupriyanov and Lukichev [27] for interface between two superconductors.

The first condition on the Usadel equations ensures continuity of the
supercurrent flowing through the S/F boundary at any value of the interfacial
transparency. Going over to the modified Usadel functions $\Phi_{S}$ and
$\Phi_{F}$, the first boundary condition has the form:
\begin{equation}
\left.  \frac{1}{\tilde{\omega}}\gamma\xi G_{F}{}^{2}\Phi_{F}^{\prime
}\right\vert _{x=0}=\left.  \frac{1}{\omega}\xi_{S}G_{S}{}^{2}\Phi_{S}%
^{\prime}\right\vert _{x=0},
\end{equation}
Here $\gamma=\rho_{S}\xi_{S}/\rho_{F}\xi$ is the proximity effect parameter,
which characterizes the intensity of superconducting correlations induced in
the F layer, and vice versa, an intensity of magnetic correlation induced into
the S layer; $\rho_{S,F}$ are the resistivities of the metals in the normal state.

The boundary condition (4) takes into account the effect of quasiparticle DOS
of the metals in contact. The second relation takes into consideration the
effects of a finite transparency (electrical quality) of the interface. For
$\Phi(\omega,x)$ - parametrization the second boundary condition becomes
\begin{equation}
\xi\gamma_{BF}G_{F}\Phi_{F}^{\prime}|_{x=0}=\tilde{\omega}G_{S}(\Phi
_{S}/\omega-\Phi_{F}/\tilde{\omega})|_{x=0},
\end{equation}
where\ $\gamma_{BF}$\ is the parameter that characterizes the effects of a
finite transparency of the interface. For $\gamma_{BF}=0$, i.e., for a fully
transparent boundary, condition (5) goes over to $\Phi_{S}/\omega=\Phi
_{F}/\widetilde{\omega}$. The expression for $\gamma_{BF}$ can be written
through more convenient values: $\gamma_{BF}=R_{B}/\rho_{F}\xi$, where $R_{B}$
is the product of the S/F boundary resistance and its area [27].

The relations (4) and (5) generalize the proximity effect problem with an
arbitrary interface transparency for the case of a normal metal with
ferromagnetic order. The additional physical condition we assumed\ is that the
exchange splitting of the momentum subbands, $p_{F}^{\pm}=\sqrt{2m}\sqrt
{E_{F}\pm H_{e}}$, is substantially smaller than the Fermi energy $E_{F}$ ($m$
is the effective mass of an electron). For most magnetic materials the
momentum renormalization is not so important as the frequency renormalization,
because $H_{e}>>\omega_{n}\symbol{126}\pi T_{C}$\ while $H_{e}<<E_{F}$ and due
to this\ the difference in the DOS and transparencies of the S/F interface for
electrons with opposite spin orientations can be neglected.

According to the Green functions formalism, if the functions $G_{S,F}%
(x,\omega)$\ and\ $F_{S,F}(x,\omega)$ are known, that is all we need to be
able to describe, at least in principle, any superconducting and magnetic
properties of the system. We draw attention to the feature important for
further conclusions: due to superconductivity\ these is only a single space
length - the respective superconducting coherence length, \ $\xi_{S}%
$\ or\ $\xi_{F}$, - that encounters into the differential equations (1) and
(3). So, \textit{due to superconductivity the coordinate dependences of both
superconducting and magnetic properties of each layer have the same typical
space scale}.

\subsection{Analytical solutions}

The proximity effect for an SF structure with a thin F metal, $d_{F}<<(\xi
_{F},\xi)$, can be reduced to consideration of the boundary value problem for
the S layer [2,26,28]. Indeed, the differential equation (3) can be solved by
iteration with respect to the parameter $d_{F}/\xi_{F}$ ($d_{F}/\xi$). To a
first approximation one can neglect the nongradient term and, taking into
account that $\Phi_{F}^{^{\prime}}(-d_{F})=0$, we obtain $\Phi_{F}(x)=const$.
In the next approximation in $d_{F}/\xi$ we find, after linearizing Eq.(3),
\begin{equation}
\Phi_{F}(\omega,x)=\frac{\tilde{\omega}\Phi_{F}(\omega,0)}{\xi\pi T_{C}G_{F}%
}\left(  x+d_{F}\right)
\end{equation}
Here we have again taken into account the condition that $\Phi_{F}^{^{\prime}%
}(-d_{F})=0$. Determining $\Phi_{F}^{^{\prime}}(0)$ from Eq. (6) and
substituting it into boundary conditions (4) and (5), we obtain the boundary
condition for the function $\Phi_{S}(\omega,x)$. We have (here we restore the
spin index):%
\begin{equation}
\xi_{S}G_{S}\Phi_{S}^{^{\prime}}|_{x=0}=\gamma_{M}\tilde{\omega}_{\sigma}%
\Phi_{S}\left[  \pi T_{C}\left(  1+\frac{2G_{S}\gamma_{B}\tilde{\omega
}_{\sigma}}{\pi T_{C}}+\frac{(\gamma_{B}\tilde{\omega}_{\sigma})^{2}}{(\pi
T_{C})^{2}}\right)  ^{1/2}\right]  ^{-1}|_{x=0}%
\end{equation}
where $\widetilde{\omega}_{\sigma}\equiv\omega+i\sigma H_{e}$. The unknown
value of the function $\Phi_{F}(\omega,x=0)$ is defined by the relation:
\begin{equation}
\Phi_{F}(\omega,0)=G_{S}\Phi_{S}\left[  \omega\left(  \frac{\gamma_{B}}{\pi
T_{C}}+\frac{G_{S}}{\tilde{\omega}_{\sigma}}\right)  \right]  ^{-1}|_{x=0}%
\end{equation}
We introduce the effective boundary parameters, $\gamma_{M}=\gamma d_{F}/\xi$
and $\gamma_{B}=\gamma_{BF}d_{F}/\xi$, instead of $\gamma$ and $\gamma_{B}$.
As a result, the problem of the proximity effect for a massive superconductor
with a thin ferromagnet layer reduces to solving the equations (1) and (3) for
a semi-infinite S layer with the boundary conditions (7) on the external side
and BCS type on infinity. The spatial dependence of the function $\Phi
_{F}(\omega,x)$ in the F layer can be neglected due to the mesoscopic
thickness $d_{F}<<(\xi,\xi_{F})$ of the latter; Eq. (8) determines the value
of $\Phi_{F}(\omega,0)$.

One can directly see, that via the boundary condition, Eq. (7), electronic
spin 'up' and spin 'down' subbands lost its equivalence in the S matel too.
\textit{Spin discrimination means magnetism of a metal}. The penetration of
the magnetic correlation into the superconducting layer is governed by the
proximity effect parameter $\gamma_{M}$, i.e., by the electron density of
states on contacting metals. For high quasiparticle's density in the F metal
in comparison to that in the S counterpart (a large value of $\gamma_{M}$) the
equilibrium diffusion of these quasiparticles into the superconductor leads to
an effective leakage of magnetic order into the S layer and strong suppression
of superconductivity near the S/F interface. In the opposite case, $\gamma
_{M}<<1$, the influence of the F layer on properties of the S metal is weak;
it even vanishes if $\gamma_{M}\longrightarrow0$. Opposite is the behavior of
the superconductivity on this parameter. So, to increase magnetic correlation
near the S/F interface one should increase the parameter $\gamma_{M}$; in
order to increase superconducting correlation one should decrease this
parameter. Of course, the electric quality of the interface is also important
for the penetration of magnetic and superconducting correlations from one
metal into another.

There are three parameters which enter the model: $\gamma_{M}$ is the measure
of the strength of proximity effect between the $S$ and $F$ metals,
$\gamma_{B}$ describes the electrical quality of the SF boundary, and $H_{e}$
is the energy of the exchange correlation in the $F$ layer. In a general case,
the problem needs self-consistent numerical solution. Here, to consider the
new physics we are interested in, we will not discuss the quantitative
calculations, but will use analytical ones obtained earlier in Refs.\ [26,28]
for two limits: (a) $\gamma_{M}<<1$, small strength of the proximity effect -
low suppression of the order parameter in the S layer near the S/F boundary,
and (b) $\gamma_{M}>>1$, strong suppression of the order parameter in the S
layer near the S/F boundary. Note that the results obtained are applicable to
any value of the S/F boundary transparency, as we made no specific assumption
about $\gamma_{B}$ in the derivation below.

\textit{Weak proximity effect. }If $\gamma_{M}<<1$, one can find an explicit
expression for $\Phi_{S}(\omega,x)$\ in the form
\begin{equation}
\Phi_{S}(\omega,x)=\Delta_{0}\{1-\gamma_{M}\beta\tilde{\omega}\frac
{\exp(-\beta x/\xi_{S})}{\gamma_{M}\beta\tilde{\omega}+\omega A(\omega)}\}
\end{equation}
where $\beta=[(\omega^{2}+\Delta_{0}^{2})/\pi T_{C}]^{1/2}$ and\ $A(\omega
)=\left[  1+\gamma_{B}\tilde{\omega}\left(  \gamma_{B}\tilde{\omega}%
+2\omega/\beta^{2}\right)  /(\pi T_{C})^{2}\right]  ^{1/2}$. As one can
expect, the magnetic correlation spreads into the S film over a distance\ of
about $\xi_{S}$ and it can much exceed the distance of the superconducting
correlation spreading into the F film $\xi_{F}$. If H$_{e}\longrightarrow0$
(i.e., $\tilde{\omega}\longrightarrow\omega$) the result (9) restores with
that for the SN bilayer in the limit in question (see, e.g., Ref. [2]). For
the function $\Phi_{F}(\omega,0)$ we obtain
\[
\Phi_{F}(\omega,0)=\Delta_{0}\tilde{\omega}_{S}/(\gamma_{B}\tilde{\omega}%
\beta^{2}+\omega)
\]

\textit{Strong proximity effect}. When $\gamma_{M}>>1$, the behavior of
$\Phi_{S}(\omega,x)$ near the S/F boundary, $0<x<<\xi_{S}$, is given by%
\begin{equation}
\Phi_{S}(\omega,0)=B(T)\{(\pi T_{C}+\gamma_{B}\tilde{\omega})/\gamma_{M}%
\tilde{\omega}\}
\end{equation}
Here $B(T)=2T_{C}[1-(T/T_{C})^{2}][7\zeta(3)]^{-1/2}$ (see Ref. [2]) and
$\zeta(3)\cong1.2$ is the Riemann $\zeta$ function. The function $\Phi
_{F}(\omega,0)$ in this approximation is read%
\[
\Phi_{F}(\omega,0)=B(T)\pi T_{C}/\gamma_{M}\omega
\]
It is seen that the proximity-induced superconductivity in the F layer is
independent of the boundary transparency, but decreases with increase of
$\gamma_{M}$. To obtain the results for larger distance, $x\gtrsim\xi_{S}$,
the equations should be solved numerically by a self-consistent procedure. We
will not discuss these results here.

\section{Magnetic proximity effect manifestation}

An important feature of the results obtained for the SF structure is that the
modified Usadel function for the S layer $\Phi_{S}(\omega,x)$, Eqs. (9) and
(10), directly depends on the exchange field of the F metal. That is the
reason to speak about the exchange correlation that has been induced into the
S layer due to superconductivity. In this section we discuss a few examples of
such 'magnetic proximity effect' manifestation.

\subsection{Phase variation at SF interface}

Comparing the results for an SF structure with those for an SN bilayer, one
can find a fundamental aspect, that leads to new physical consequences;
namely, the $\Phi_{S}(\omega,x)$ is a complex function near the S/F interface.
As a result, the additional 'superconducting phase rotation' (a phase jump\ on
the S/F interface for our approximation of a thin ferromagnetic layer
$d_{F}<<\xi_{F}$) occurs at the S/F interface. To illustrate this, let us
take, for simplicity, a\ structure with favorable for magnetic effects
interface parameters: $\gamma_{M}>>1$ and $\gamma_{B}=0$. Then, as follows
from Eq. (10), the modified Usadel function at S/F interface $\Phi_{S}%
(\omega,0)$ can be written in the form
\begin{equation}
\Phi_{S}(\omega,0)=B(T)(\pi T_{C}/\gamma_{M})\frac{\exp(-i\theta)}{(\omega
^{2}+H_{e}^{2})^{1/2}}\text{,}%
\end{equation}
with $\theta=\arctan(H_{e}/\omega)$.\ Taking into account that a typical value
of $\omega\symbol{126}\pi T_{C}$, one can see that in the limit $H_{e}>>\pi
T_{C}$ the correlation function acquires an additional $\pm\pi/2$ phase shift
in comparison with the similar function for the SN bilayer. For an SF
multilayred system with strong enough ferromagnetism of the layers the phase
shift can be summarized or subtracted, depending on mutual orientation of F
layers magnetizations, leading to new effects in superonductivity of SF
hybride structures. Namely, one can show, that proximity induced magnetizm of
the S layers makes preferrable the $\pi$-phase superconductivity of the system
for parallel directions of the exchange fields; for antiparallel
magnetizations orientation and low temperature, the critical current can be
even enhanced [12-15].

\subsection{Suppression of the superconducting order parameter by an exchange
field}

Another feature of the S/F boundary is that the gap $\Delta_{S}(x)$ is
suppressed near the interface more strongly than in the SN case. This is not
surprising, since one would expect that induced ferromagnetism suppresses the
superconducting order parameter at some distance into the S layer in excess of
that for nonmagnetic normal layer. Suppression increases with the increase of
the exchange energy $H_{e}$\ and of electrical quality of the interface; far
from the interface, $x>>\xi_{S}$, the bulk superconductivity is restored.

Using $\Phi_{S}(\omega,x)$ (9) and the self-consistency condition (2) one can
find the spatial variation of the order parameter in the $S$ layer $\Delta
_{S}(x)$ for different values of $\gamma_{B}$ and $\gamma_{M}<<1$. The
exchange interaction influence on the spatial variation of the order parameter
in the S layer is shown in Fig. 1. Namely, the dependence of difference of the
order parameters for the case when magnetic interaction is turned off (i.e.,
an SN\ bilayer) and with ferromagnetic correlation (a SF bilayer) as function
of distance from the interface is shown; the boundary parameters, $\gamma_{M}$
and $\gamma_{B}$,\ are fixed. It is seen, that influence of magnetism
decreases with increasing the distance from the S/F boundary. The scale at
which superconductivity reaches the value for a SN\ bilayer is $\xi_{S}$ from
the interface. The curves in Fig. 1 illustrate the spatial dependencies
of\ the induced exchange correlation in the superconductor for the case of
vanishing interface resistance $\gamma_{B}=0$. With an increase of the SF
boundary resistance the electrical coupling of the\ S\ and F\ metals decreases
and in the limit $\gamma_{B}\longrightarrow\infty$ the metals become decoupled.

\subsection{Exchange field spin-splitting of DOS and intra-gap states}

Spin splitting of DOS and intergap states in the S layer are other
manifestations of magnetic correlation leakage\ into a superconductor. Note
that the magnetic layer does not influence the DOS of the normal metal. In
this case the decay length is extremely small $\sim p_{F}^{-1}\simeq
1\mathring{A}$ and the effect can be neglected.

The Green functions for the S layer $G_{S\uparrow\uparrow}(\omega,x)$ and
$G_{S\downarrow\downarrow}(\omega,x)$ for both spin subbands can be obtained
using solutions for the functions $\Phi_{S}(\omega,x)$ with $\tilde{\omega
}=\omega+iH_{e}$\ and $\tilde{\omega}=\omega-iH_{e}$, respectively. Performing
the analytical continuation to the complex plane by the substitution
$\omega\rightarrow-i\varepsilon$ we calculate the spatial dependence of
quasiparticle DOS for spin 'up' and 'down' subband: $N_{S\uparrow}%
(\varepsilon,x)=ReG_{S\uparrow\uparrow}(\omega,x)$ and $N_{S\downarrow
}(\varepsilon,x)=ReG_{S\downarrow\downarrow}(\omega,x)$, respectively The
total density of states for quasiparticles, by definition, is given by
$N_{S}(\varepsilon,x)=N_{S\uparrow}(\varepsilon,x)+N_{S\downarrow}%
(\varepsilon,x)$. Using Eqs. (9) and (10), one can obtain the explicit
expressions for the total DOS, as well as for the specified spin subband. The
resulting expressions, which are cumbersome to be presented here, imply that
for $H_{e}\neq0$, $\gamma_{M}\neq0$, and $\gamma_{B}\neq0$ the density of
quasiparticle states is spin-splitted: $N_{S\uparrow}(\varepsilon,x)\neq
N_{S\downarrow}(\varepsilon,x)$. This is because of the initial exchange field
splitting of the Fermi surface in the F metal, which is manifested in the
characteristics of the united system --- the SF bilayer. The symmetry of the
density of states with respect to the energy variable is also lost:
$N_{S\uparrow}(\varepsilon>0,x)\neq N_{S\uparrow}(\varepsilon<0,x)$. However,
as one can expect from the fermionic symmetry, the spin-up particles and
spin-down holes have the same DOS, and likewise for spin-down particles and
spin-up holes; as a result the total density $N_{S}(\varepsilon,x)$ is
symmetric: $N_{S}(\varepsilon>0,x)=N_{S}(\varepsilon<0,x)$.

In Fig. 2 representative $N_{S\uparrow}(\varepsilon,x)$ dependences at
different distances from the S/F interface are presented for $H_{e}=5\pi
T_{C}$ and $\gamma_{M}=0.1$, and vanishing boundary resistance ($\gamma_{B}%
=0$). In Fig. 3 the same dependence is presented for $x/\xi_{S}=1$ and
different values of the exchange energy. We find that all features mentioned
above are saved on a distance of a scale $\xi_{S}$ from the SF boundary. The
spin-splitting decreases with an increase of the distance from the boundary
and vanishes in the bulk of the$\ $S layer (see curve 4 in Fig. 2).

Other important features, shown in Figs. 2 and 3, are the local states that
appear inside the energy gap at the distances up of a few $\xi_{S}$ from the
S/F boundary. These intergap states are absent far from the S/F\ interface,
and also if $H_{e}=0$. For small values of $\gamma_{M}$ and $\gamma_{B}=0$, as
follows from the expression (9), $N_{S\uparrow}(\varepsilon,x)$ has
singularity for
\begin{equation}
\varepsilon=\pm\Delta_{0}\{1-\frac{\gamma_{M}\beta_{\varepsilon}%
\tilde{\varepsilon}}{\varepsilon+\gamma_{M}\beta_{\varepsilon}\tilde
{\varepsilon}}\exp(-\beta_{\varepsilon}x/\xi_{S})\}
\end{equation}
where $\beta_{\varepsilon}^{2}=(\Delta_{0}^{2}-\varepsilon^{2})^{1/2}/\pi
T_{C}$ and$\ \tilde{\varepsilon}=\varepsilon-H_{e}$. We found the singularity
inside the superconducting gap, $-\Delta_{0}<\varepsilon<\Delta_{0}$, by
numerical calculations [19]. The local states are definitely not due to the
spatial variation of the pair potential, but due to Cooper pairs breaking in
the superconductor by the exchange-induced magnetic correlation. The region of
their existence increases with the increasing of $H_{e}$, or increasing pair
breaking effects. In the absence of spin-flip (e.g., spin-orbit) scattering,
the subgap bands accommodate quasiparticles with a definite ('up' or 'down')
spin direction. These bands bear superficial resemblance to both the bands
observed at interface of superconductor and perfectly insulating ferromagnet
[29] and bulk superconductor containing finite concentrations of magnetic
impurities [30,31].

\subsection{Induced magnetization of the S layer}

As we saw above, the influence of the ferromagnet on the superconductor is
reflected in a nonzero value of the difference in the DOS for spin-up and
spin-down unpaired electrons, $N_{S\uparrow}(\varepsilon,x)$ and
$N_{S\downarrow}(\varepsilon,x)$. This DOS splitting causes an effective
magnetization $M_{S}(x)$ of the S layer, that can be found using the
relation:
\begin{equation}
M_{S}(x)/M_{O}=\int_{0}^{\infty}d\varepsilon\{N_{S\uparrow}(\varepsilon
,x)-N_{S\downarrow}(\varepsilon,x)\}f(\varepsilon)
\end{equation}
where $M_{O}=gS_{e}\mu_{B}(=\mu_{B})$\ is a quasiparticle magnetic
moment,\ $S_{e}=1/2,g=2$ and $f(\varepsilon)=1/[\exp(\varepsilon/T)+1]$ is the
Fermi distribution function. Figure 4 illustrates the mechanism of proximity
induced magnetization of the S layer. For $T<T_{C}$ we took $f(\varepsilon
)=1$, i.e., all states below Fermi level are filled (dashed regions in Fig.
4), while all states above Fermi energy are empty. One can directly see from
the figure that the S layer acquires a nonzero magnetic moment. This
suggestion is confirmed by numerical calculations of $M_{S}(x)$ Eq.(13) shown
in Figs. 5, 6. Figure 5 shows the magnetization of the superconductor versus
distance from the S/F interface for fixed boundary parameters. The same
magnetic characteristics but for a SF\ sandwich with fixed exchange energy and
boundary transparency, and different proximity effect strength are presented
in Fig. 6.\ 

\subsection{Experiment}

There are only a few experimental reports devoted to the questions discussed
here. Interplay between magnetism and superconductivity in Nb/Co multilayers
was recently investigated by Ogrin \textit{et al}. [32]. The upper critical
fields of the samples were measured for the field applied parallel to the
plane, H$_{C_{2}||}$ and perpendicular to the plane H$_{C_{2}\perp}$ of the
films. Effective thickness of the Co layer, d$_{eff}$ , they define through
the well known relation:%
\[
d_{eff}=%
%TCIMACRO{\QOVERD{(}{)}{\Phi_{0}}{2\pi H_{C_{2}\perp}}}%
%BeginExpansion
\genfrac{(}{)}{}{}{\Phi_{0}}{2\pi H_{C_{2}\perp}}%
%EndExpansion
^{1/2}\frac{H_{C_{2}\perp}}{H_{C_{2}||}}%
\]
where $\Phi_{0}$ stands for flux quantum. Experiments revealed that the
effective thickness of the magnetic layer in $Nb/Co$ structures\ is usually
much larger than its physical thickness $d_{Co}$. For example, taking the data
on sample with d$_{Co}$ = 1.8.nm, the authors obtained a value $d_{eff}=$ 12
nm, so that $d_{eff}>>d_{Co}$. The 'increase' of the thicknesses was so great
that in all samples, except for those with extremely thin magnetic layers, the
crossover to a 3D state superconductivity is never in fact observed
experimentally. This is to be contrasted with the case of nonmagnetic spacer
layers, where these two length scales are comparable. Taking into account our
results, we explain the rise of the effective magnetic layer thickness in the
$Nb/Co$ multilayer as an impact of proximity effect. Namely, the induced
magnetic correlation into the $S$ layer depletes Cooper pairs density at the
$SF$ boundary, which results in an excess thickness of the magnetic layer.

The modification of the DOS in mesoscopic superconducting strips of $Al$ under
the influence of magnetic proximity effect of a classical ferromagnet $Ni$ has
also been studied both theoretically and experimentally\ in [33]. However,
since the tunnel spectroscopy experiments were carried out with a nonmagnetic
probe, the authors could not measure spin-denendent local DOS in the
superconducting side.

The interest in the magnetic proximity effect has been increased with the
development of experimental techniques like neutron reflectometry and muon
spin rotation, which allow to determine accurately the spatial distribution of
magnetic moments. For example, very recently multilayered system YBa$_{2}%
$Cu$_{3}$O$_{7}$/La$_{2/3}$Ca$_{1/3}$MnO$_{3}$ have been studied by neutron
reflectometry in [34]. Evidence for a characteristic difference between the
structural and magnetic depth profiles is obtained from the occurrence of a
structurally forbidden Bragg peak is a ferromagnetic state. The authors
discussed findings in two possible scenarios: a sizable magnetic moment within
the Slayer antiparallel to one in the F layer (inverse proximity effect), or a
"dead" region in the F layer with zero net magnetic moment.

\section{Conclusion}

In recent years, advances in materials growth and fabrication techniques have
made it possible to create heterostructures with high quality interfaces.
Taking into account that ferromagnet-superconductor hybrid systems have great
scientific importance, and are promising for application in spin-electronics,
it is not surprising that interest to these hybrid materials has been renewed.
As far as the thickness of superconducting and magnetic metals in such
structures may be a few atomic periods, understanding of how the proximity
effects modify electronic properties of S/F interfaces is growing in importance.

We have studied in the magnetic correlations acquired by a superconductor at
S/F interface due to a proximity effect. We have found that an equilibrium
exchange of electrons between the F and S metals results not only in proximity
induced superconductivity of the F metal, as was found earlier, but in
proximity induced magnetism of the S metal, too. The magnetic correlations
spread over a large distance which is of the order of the superconducting
coherence length $\xi_{S}$\ and can exceed both the ferromagnetic and the
superconducting films thicknesses. That is why the existence of these magnetic
properties of the S metal is quite important for SF nanoscale structures and
should be taken into account while comparing theoretical results with
experimental data. Summarizing the results, we should stress that for SF
nanoscopic hybrid structures both phenomena, -- the superconducting and the
magnetic proximity effects, -- take place simultaneously, and both should be
paid attention to.

\emph{Acknowledgements}. We wish to dedicate this paper to V.G. Bar`yakhtar,
our Master who played a significant, exceptional role in our post-student
life, on the occasion of his 75th birthday, and to wish him continuing health
and vigour in pursuing his scientific interest. The authors would like to
thank V. V. Ryazanov, A.I. Buzdin, L. Tagirov, and M. A. Belogolovskii for
valuable discussions of some questions of proximity effect phenomena. We also
acknowledge E. A.\ Koshina for performing numerical calculations.

\newpage

1. \textit{Wolf E. L.} Principles of electron tunneling spectroscopy. -
Oxford: University Press, 1985.

2. \textit{Golubov A. A., Kupriyanov M. Yu., Il'ichev E}. //Rev. Mod. Phys. -
2004.- \textbf{76}, \#2.- p.411-469.

3. \textit{Izumov Yu. A., Proshin Yu. N., Khusainov M. G.} //Usp. Fiz. Nauk. -
2002.- \textbf{45}, \#2.- p.114-154.

4. \textit{Buzdin A. I., Bulaevskii L. N., Panyukov S. V.} //Sov. Phys. JETP
Lett. -1982.- \textbf{35}, \#4. - p.147-148.

5. \textit{Buzdin A. I., Kupriyanov M. Yu.\ }//Sov. Phys. JETP Lett. - 1991.-
\textbf{53}, \#6.- p.308-312.

6. \textit{Ryazanov V. V., Oboznov V. A., Rusanov A. Yu., Veretennikov A. V.,
Golubov A.\ A., Aarts J.} //Phys. Rev. Lett. -2001.- \textbf{86}, \#11-12.- p.2427-2430.

7. \textit{Ryazanov V. V., Oboznov V. A., Veretennikov A. V., Rusanov A. Yu.}
//Phys. Rev. B - 2002 - \textbf{65}, \#2. - 020501-4(R).

8. \textit{Blum Y., Tsukernik A., Karpovski M., Palevski A.} //Phys. Rev.
Lett. - 2002. - \textbf{89}, \#18.- p.187004-4.

9. \textit{Kontos T.,Aprili M., Lesueur J., Genet F., Stephanidis B., Boursier
R}. //Phys. Rev. Lett. - 2002.- \textbf{89}, \#13. - 137007-4.

10. \textit{Kontos T., Aprili M., Lesueur J., Grison X.} //Phys. Rev. Lett. -
2001.- \textbf{86}, \#2. - p.304-307.

11. \textit{Guichard W., Aprili M., Bourgeois O., Kontos T., Lesueur J.,
Gandit P.\ }//Phys. Rev. Lett. - 2003.- \textbf{90}, \#16.- p.167001-4.

12. \textit{Bergeret F. S., Volkov A. F., Efetov K. B.} //Phys. Rev. Lett. -
2001.- \textbf{86}, \#14.- p.3140-3143.

13. \textit{Krivoruchko V. N., Koshina E. A.} //Phys. Rev. B - 2001.-
\textbf{64}, \#17.- p.172511-4.

14. \textit{Golubov A. A., Kupriyanov M. Yu., Fominov Ya. V.} //JETP Lett. -
2002.- \textbf{75}, \#11.- p.709-713.

15. \textit{Koshina E. A., Krivoruchko V. N. }//Low. Temp. Phys. - 2003.-
\textbf{29}, \#8.- p.642-649.

16. \textit{Halterman K.,Valls O. T.} //Phys. Rev. B -2002.- \textbf{65},
\#1.- p.014509-12.

17. \textit{Fazio R., Lucheroni C.} //Europhys. Lett. - 1999.-\ \textbf{45},
\#6. - p.707-713.

18. \textit{Halterman K., Valls O. T.} //Phys. Rev. B - 2004.- \textbf{69},
\#1.- 014517-11.

19. \textit{Krivoruchko V. N., Koshina E. A.} //Phys. Rev. B - 2002-
\textbf{66}, \#1.- 014521-6.

20. \textit{Bergeret F. S., Volkov A. F., Efetov K. B.}\ //Phys. Rev.B -2004.-
\textbf{69}, \#17.- p.174504-5.

21. \textit{Millis A., Rainer D., Sauls J. A.} //Phys. Rev. B - 1988.-
\textbf{38,} \#7.- p.4504-4515.

22. \textit{Tokuyasu T., Sauls J. A., Rainer D.} //Phys. Rev. B - 1988.-
\textbf{38}, \#13.- 8823-8833.

23. \textit{Fogelstr\"{o}m M. }//Phys. Rev. B - 2000.- \textbf{62}, \#17.- p.11812-11819.

24. \textit{Usadel K. D. }//Phys. Rev. Lett. - 1970.- \textbf{25}, \#8.- 507-509..

25. \textit{A. M. Swidzinski\"{\i}}. Spatially Nonuniform Problems of the
Theory of Superconductivity. - Moscow: Nauka, 1982.

26. \textit{Koshina E. A., Krivoruchko V. N.}//Low. Tem. Phys. - 2000. -
\textbf{26}, \#2.- p.115-120.

27. \textit{Kuprijanov M. Yu., Lukichev V. F. }//Sov. Phys. JETP - 1988.-
\textbf{94}, \#6.- p.139-149.

28. \textit{Koshina E. A., Krivoruchko V. N.//}Phys. Rev. B - 2001.-
\textbf{63}, \#22.- 224515-8.

29. \textit{DeWeert M. J., Arnold G. B.} //Phys. Rev. B - 1989.- \textbf{39},
\#16.- 11307-11319.

30. \textit{Shiba H. }//Prog. Theor. Phys. - 1968.- \textbf{40}, \#3.- p.435-451.

31. \textit{Rusinov A.J. }//Sov. Phys. JETP Lett. - 1969.- \textbf{9,}\#1.- p.85-89.

32. \textit{Ogrin F. Y., Lee S. L., Hillier A. D., Mitchell A., Shen T.-H.
}//Phys. Rev. B - 2000.- \textbf{62}, \#9.- 6021-6026.

33.\textit{ Sillanp\"{a}\"{a} M. A., Heikkil\"{a} T. T., Lindell R. K.,
Hakonen P. J.} //Europhys. Lett. - 2001.- \textbf{56}, \#4.- p.590-595.

34. \textit{Stahn J., Chakhalian J., Niedermayer Ch., Hoppler J., Gutberlet
T., Voigt J., Treubel F., Habermeier H-U., Cristiani G., Keimer B., Bernhard
C}. //Phys. Rev. B. -2005.- \textbf{71},\#14 - p.140509-4(R).

\newpage

\begin{center}
\textbf{Figure Captures}
\end{center}

Fig. 1. The difference of the superconducting order parameter in the S layer
versus distance from the interface for SN and SF structures with the same
boundary parameters ($\gamma_{M}=0.1$, $\gamma_{B}=0$), and different
ferromagnetic field energy $H_{e}/\pi T_{C}=$ 8, 9, 10, 12 and 15.

Fig. 2. Normalized density of state for spin 'up' quasiparticles in the S
layer of the SF sandwich for $\gamma_{M}=0.1$ , $\gamma_{B}=0$ and $H_{e}=5\pi
T_{C}$, and various distances from the S/F interface: $x/\xi_{S}=$ 0, 1, 5,
and 30 (curves 1, 2, 3, and 4, respectively).

Fig. 3. Same as in Fig. 4 for $\gamma_{M}=0.1$, $\gamma_{B}=0.1$ and
$x=\xi_{S}$, and various ferromagnetic field energies: $H_{e}/\pi T_{C}=$ 1,
2, and 5 (curves 1, 2, and 3, respectively).

Fig. 4. Quasiparticle density of states in the S layer near the S/F interface;
$\gamma_{M}=0.1$, $\gamma_{B}=0.0$, $x=\xi_{S}$, and $H_{e}=5\pi T_{C}$. All
states above Fermi energy are empty; all states below Firmi level are filled
(dashed regions in figure).

Fig. 5. Leakage of magnetization into the S material versus distance from the
interface for SF sandwich for $\gamma_{M}=0.1$ , $\gamma_{B}=0$ , and
different exchange energies $H_{e}/\pi T_{C}=$7, 5, and 3 (curves 1, 2, and 3, respectively).

Fig. 6. Same as in Fig. 6 for $\gamma_{B}=0$ , $H_{e}=3.5\pi T_{C}$ and
different proximity effect strength $\gamma_{M}$ =\ 0.1, 0.15, 0.2 (curves 1,
2, and 3, respectively).

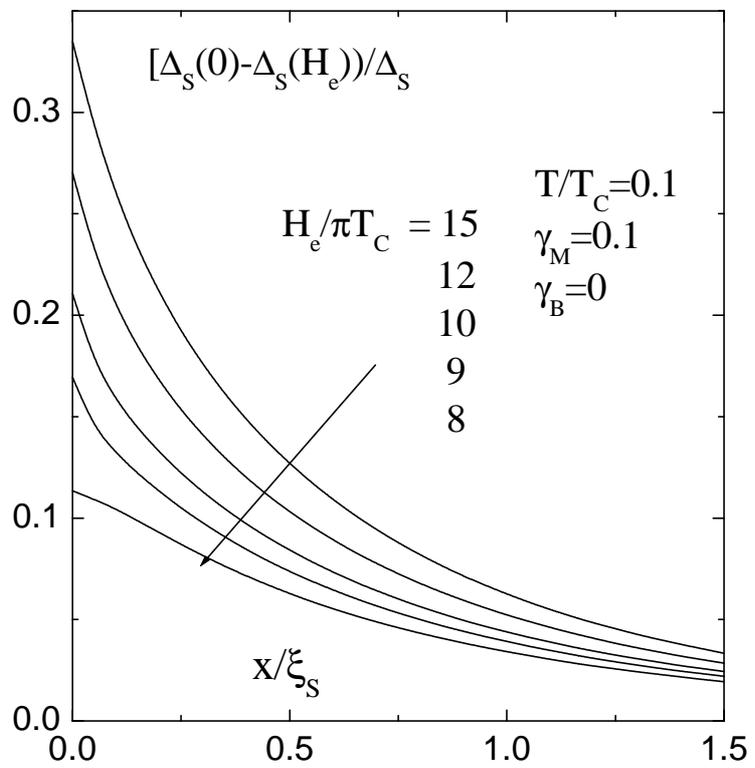
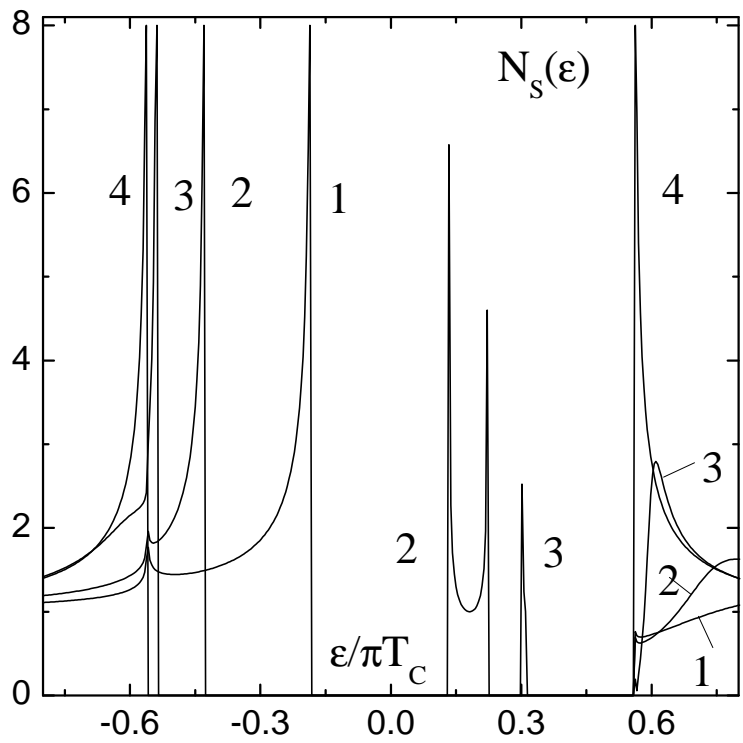
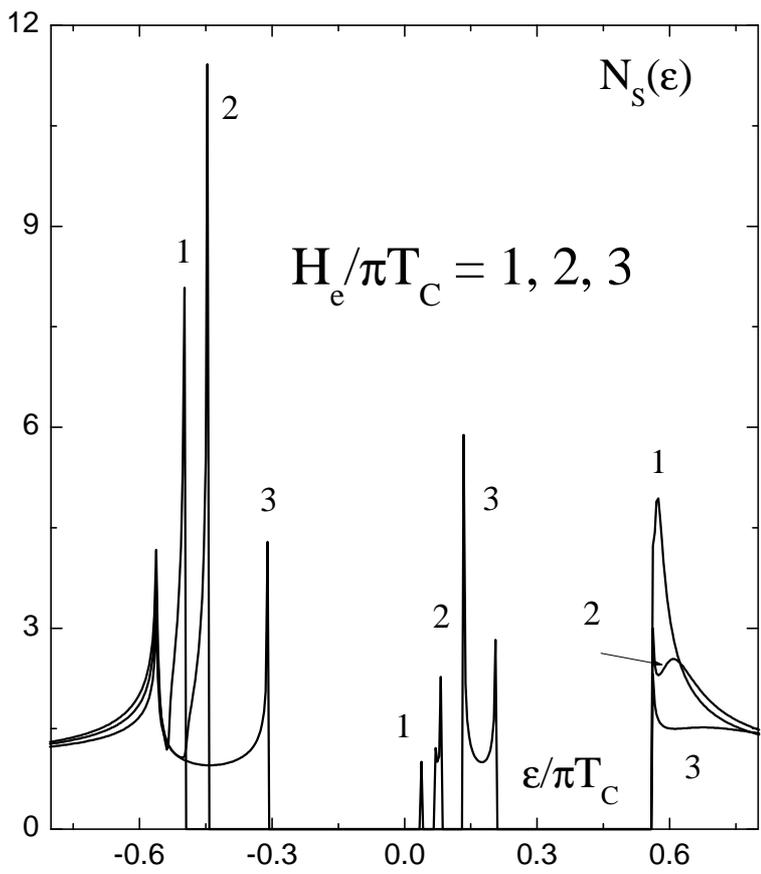
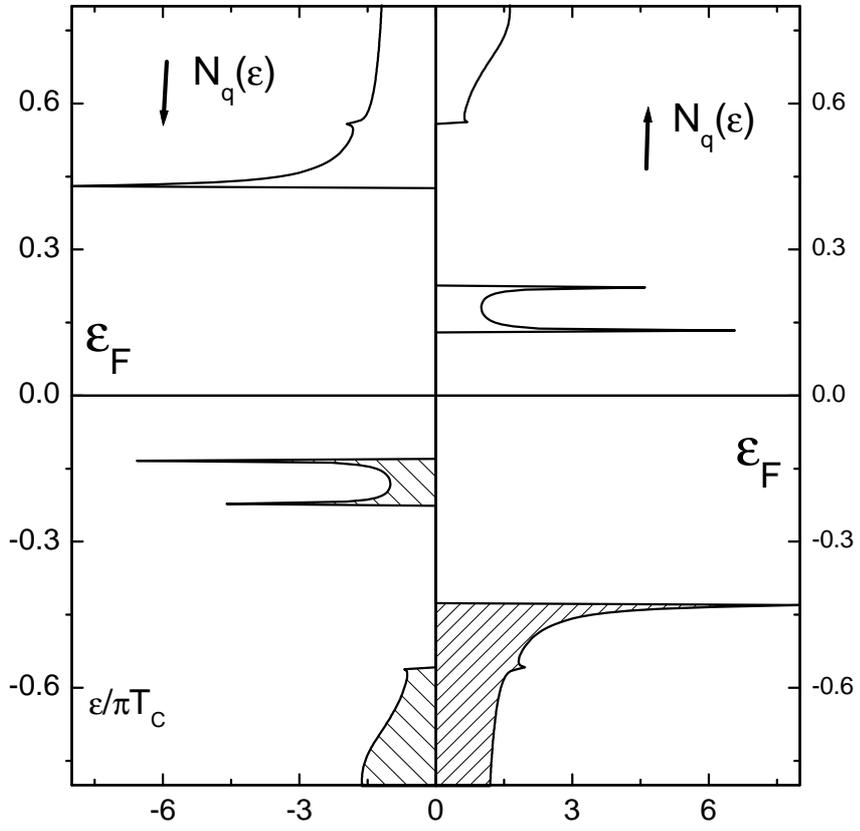
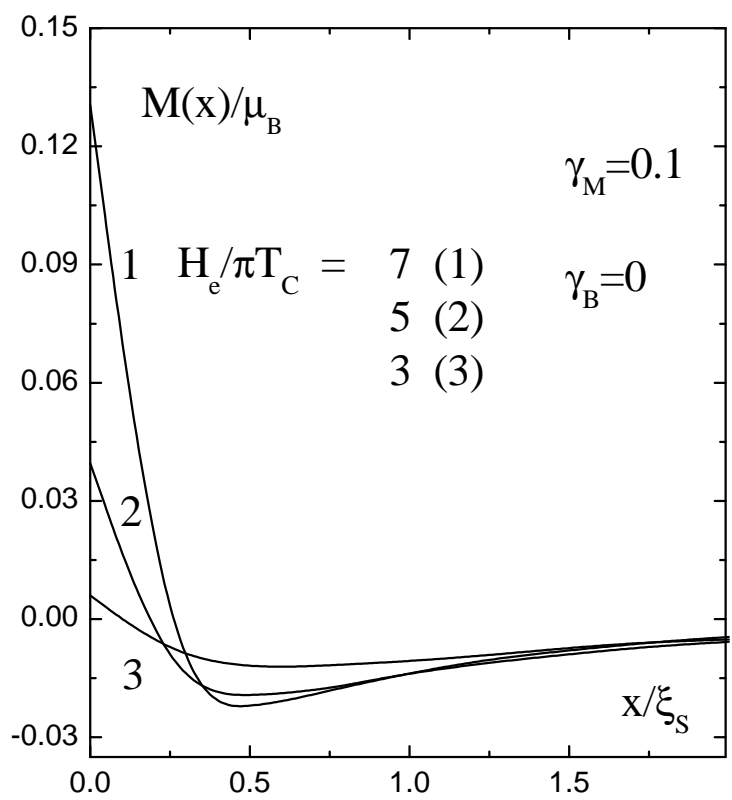
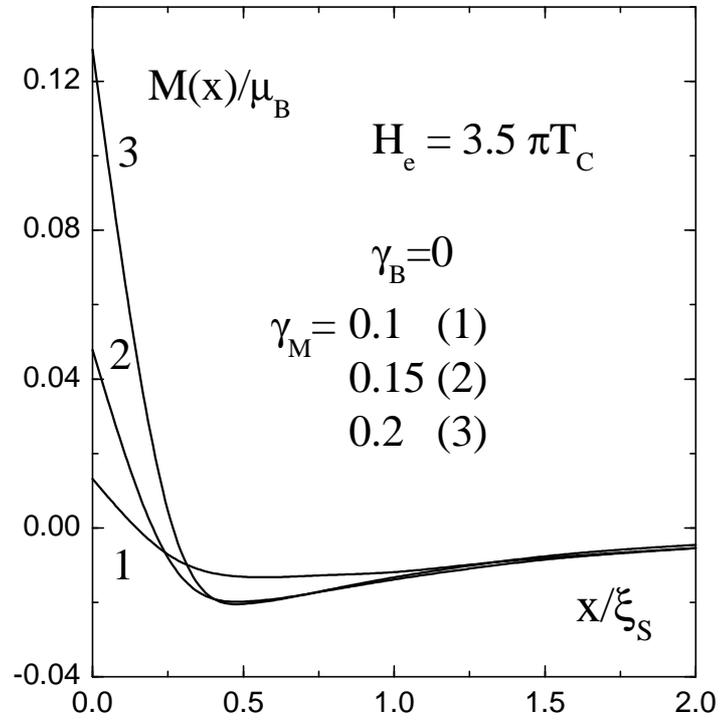
\end{document}